\documentclass[]{aa}
\usepackage{natbib}         
\usepackage{graphicx}
\usepackage{longtable}
\usepackage[switch]{lineno} 
\bibpunct{(}{)}{;}{a}{}{,} 

\def\Kepler{\emph{Kepler}}

\def\Dnu{\Delta\nu}
\newcommand\Tg{\Delta\Pi_1}

\newcommand{\HRsismo}{\Tg$  -- $\Dnu}

\def\Mmax{1.5}

\begin{document}

\title{Mixed modes in red giants: a window on stellar evolution\thanks{Table 1
is only available in electronic form at the CDS
via anonymous ftp to \texttt{cdsarc.u-strasbg.fr (130.79.128.5)}
or via \texttt{http://cdsweb.u-strasbg.fr/cgi-bin/qcat?J/A+A/}}}
\titlerunning{Stellar evolution}
\author{%
 B. Mosser\inst{1} \and
 O. Benomar\inst{2,5} \and
 K. Belkacem\inst{1} \and
 M.J. Goupil\inst{1} \and
 N. Lagarde\inst{3,9} \and
 E. Michel\inst{1} \and
 Y. Lebreton\inst{4,11} \and
 D. Stello\inst{5,9} \and
 M. Vrard\inst{1} \and
 C. Barban\inst{1} \and
 T.R. Bedding\inst{5,9} \and
 S. Deheuvels\inst{6} \and
 W.J. Chaplin\inst{3,9} \and
 J. De Ridder\inst{7} \and
 Y. Elsworth\inst{3,9} \and
 J. Montalban,
 A. Noels\inst{8} \and
 R.M. Ouazzani\inst{9} \and
 R. Samadi\inst{1} \and
 T.R. White\inst{10} \and
 H. Kjeldsen\inst{9}
 }

\institute{
 LESIA, CNRS, Universit\'e Pierre et Marie Curie,
 Universit\'e Denis Diderot, Observatoire de Paris, 92195 Meudon
 cedex, France; \texttt{benoit.mosser@obspm.fr}
 \and Department of Astronomy, The University of Tokyo, 113-033, Japan
 \and School of Physics and Astronomy, University of Birmingham,  Edgbaston, Birmingham B15 2TT, United Kingdom
 \and GEPI, CNRS, Universit\'e Denis Diderot, Observatoire de Paris, 92195 Meudon cedex,
 \and Sydney Institute for Astronomy, School of Physics, University of Sydney, NSW 2006, Australia
 \and Universit\'e de Toulouse; UPS-OMP; CNRS; IRAP; 14, avenue Edouard Belin, F-31400 Toulouse, France
 \and Instituut voor Sterrenkunde, KU Leuven, Celestijnenlaan 200D, 3001 Leuven, Belgium
 \and Institut d'Astrophysique et G\'eophysique de l'Universit\'e de Li\`ege, All\'ee du six Ao\^ut,
   17 B-4000 Li\`ege, Belgium
 \and Stellar Astrophysics Centre, Department of Physics and Astronomy,  Aarhus University, Ny Munkegade 120, DK-8000 Aarhus C, Denmark
 \and Institut f\"ur Astrophysik, Georg-August-Universit\"at G\"ottingen, Friedrich-Hund-Platz 1, 37077  G\"ottingen, Germany
 \and Institut de Physique de Rennes, Universit\'e de Rennes 1,
CNRS (UMR 6251), 35042 Rennes, France
 }


\abstract{The detection of oscillations with a mixed character in
subgiants and red giants allows us to probe the physical
conditions in their cores. }
{With these mixed modes, we aim at determining seismic markers of
stellar evolution.}
{\Kepler\ asteroseismic data were selected to map various
evolutionary stages and stellar masses. Seismic evolutionary
tracks were then drawn with the combination of the frequency and
period spacings. }
{We measured the asymptotic period spacing for more than 1170
stars at various evolutionary stages. This allows us to monitor
stellar evolution from the main sequence to the asymptotic giant
branch and draw seismic evolutionary tracks. We present clear
quantified asteroseismic definitions that characterize the change
in the evolutionary stages, in particular the transition from the
subgiant stage to the early red giant branch, and the end of the
horizontal branch.}
{The seismic information is so precise that clear conclusions can
be drawn independently of evolution models. The quantitative
seismic information can now be used for stellar modeling,
especially for studying the energy transport in the helium-burning
core or for specifying the inner properties of stars entering the
red or asymptotic giant branches. Modeling will also allow us to
study stars that are identified to be in the helium-subflash stage, high-mass
stars either arriving or quitting the secondary clump, or stars
that could be in the blue-loop stage.}


\maketitle

\section{Introduction}

High-precision photometry has revealed that red giant stars
oscillate like the Sun \citep{2009Natur.459..398D}. Unlike the
Sun, where the oscillations are pressure modes, red giants also
show gravity modes. These oscillations have been used to
distinguish between stars that are burning only hydrogen in a thin
shell around their cores and those that are additionally burning
helium inside their cores
\citep{2011Natur.471..608B,2011A&A...532A..86M,2013ApJ...765L..41S}.
The spacings between oscillation periods reported by these studies
are  significantly offset compared to their theoretical
counterparts, however, so they cannot be used for identifying specific
evolutionary tracks. However, exact measurements of the asymptotic
period spacings are now available \citep{2012A&A...540A.143M},
which are directly related to the core size
\citep{2013ApJ...766..118M}.

The combined information on the core and on the envelope of red
giants can be obtained from observing the oscillation
mixed-mode pattern \citep{2011Sci...332..205B}.
These modes result from the coupling of acoustic waves that probe
the mostly convective stellar envelope and gravity waves that
probe the dense radiative  stellar core. They share the properties
of acoustic and gravity modes. Acoustic modes have frequencies
approximately equally spaced \citep{1980ApJS...43..469T}. The
frequency difference between consecutive radial oscillation modes,
hereafter denoted $\Dnu$, is called the large frequency separation
and depends on the mean stellar density. Unlike pressure modes,
gravity modes are equally spaced in period. For dipole modes this
spacing is denoted $\Tg$ and is dependent upon the density
stratification in central regions \citep{1980ApJS...43..469T}.
Determining $\Tg$ allows us to probe the core, to monitor its
rotation
\citep{2012Natur.481...55B,2012A&A...548A..10M,2012ApJ...756...19D,2014A&A...564A..27D},
and to investigate how angular momentum is transferred between the
stellar core and the envelope \citep{2014ApJ...788...93C}.

Here, we use such frequency and period spacings to track
evolutionary stages ranging from the end of the main sequence to
the asymptotic giant branch (AGB) in a selection of stars observed
by \Kepler. Seismic observations are precise enough to derive
model-independent conclusions.

\section{Data and methods}

The data set is composed of 38 subgiants observed by \Kepler\
during at least one month \citep{2011Sci...332..213C} and of about
12\,700 red giants observed during 44 months
\citep{2013ApJ...765L..41S}. Out of these, $2800$ were selected to map the whole range in frequency spacings and masses.
This selection precludes any population analysis, but it provides an
exhaustive view on low-mass star evolution.

 We have used an automated method for measuring
the frequency spacing of radial pressure modes and have developed
a semi-automated method for measuring the period spacing of dipole
mixed modes, both based on asymptotic expansions
\citep{2011A&A...525L...9M,2012A&A...540A.143M}. The mean accuracy
of the large separation is about 0.04\,$\mu$Hz; this translates
into a relative precision at the red clump of 1\,\%. The
measurement of $\Tg$ relies on the number of mixed modes with high
signal-to-noise ratio, which depends on the evolutionary status
\citep{2009A&A...506...57D,2012A&A...540A.143M}. The precise fit
of the oscillation spectrum must account for any rotational
splitting
\citep{2012Natur.481...55B,2012A&A...548A..10M,2013A&A...549A..75G}.

For red giants, the observed gravity-mode orders are high enough
to ensure the validity of the asymptotic expansion and a high
precision of the asymptotic global parameters. Hence, the period
spacing $\Tg$ is determined with a precision better than 2\,\% and
in many cases better than 0.5\,\%. For subgiants, the gravity-mode
orders of the few observed mixed modes are small, down to 2 in
many cases, so that the quantitative use of the asymptotic
expansion may not be accurate. However, comparison with a
different approach \citep{2013ApJ...767..158B,2014ApJ...781L..29B}
indicates agreement in the obtained values of the period spacings
to within 10\,\% for subgiants and better than 3\,\% for red
giants.

We measured $\Tg$ in 1142 red giants and 36 subgiants. Reliable
measurements are impossible in oscillation spectra with low
signal-to-noise ratio or, most often, in absence of enough
gravity-dominated mixed modes. This occurs at low $\Dnu$, when
gravity-dominated mixed modes have high inertia
\citep{2009A&A...506...57D,2014arXiv1409.6121G} so that major
difficulties occur for stars with $\Dnu<5\,\mu$Hz on the red giant
branch (RGB). In that case, measurements are possible only for
bright stars seen pole-on, when the rotational structure of the
dipole mixed modes is simple since only $m=0$ modes are visible.
The number of ambiguous cases, with a large number of mixed modes
but a poor fit of the mixed-mode oscillation pattern, is less than
0.1\,\% (the spectra of 2 stars out of 2800 that were treated
remain obscure despite a decent signal-to-noise ratio; they
certainly corresponds to blended light curves).

The stellar masses and radii were derived from the seismic scaling
relations calibrated with nine red giants and eleven subgiants
\citep{2013A&A...550A.126M}. Luminosity was derived from the
stellar radius and effective temperature, assuming the
Stefan-Boltzmann law. The precision is 15-25\,\% in mass, 5-10\,\%
in radius, 15-25\,\% in luminosity
\citep{2012ApJ...760...32H,2012ApJ...757...99S}.

\section{Seismic HR diagram}

\begin{figure*}
\bigskip\bigskip
\includegraphics[width=15.5cm]{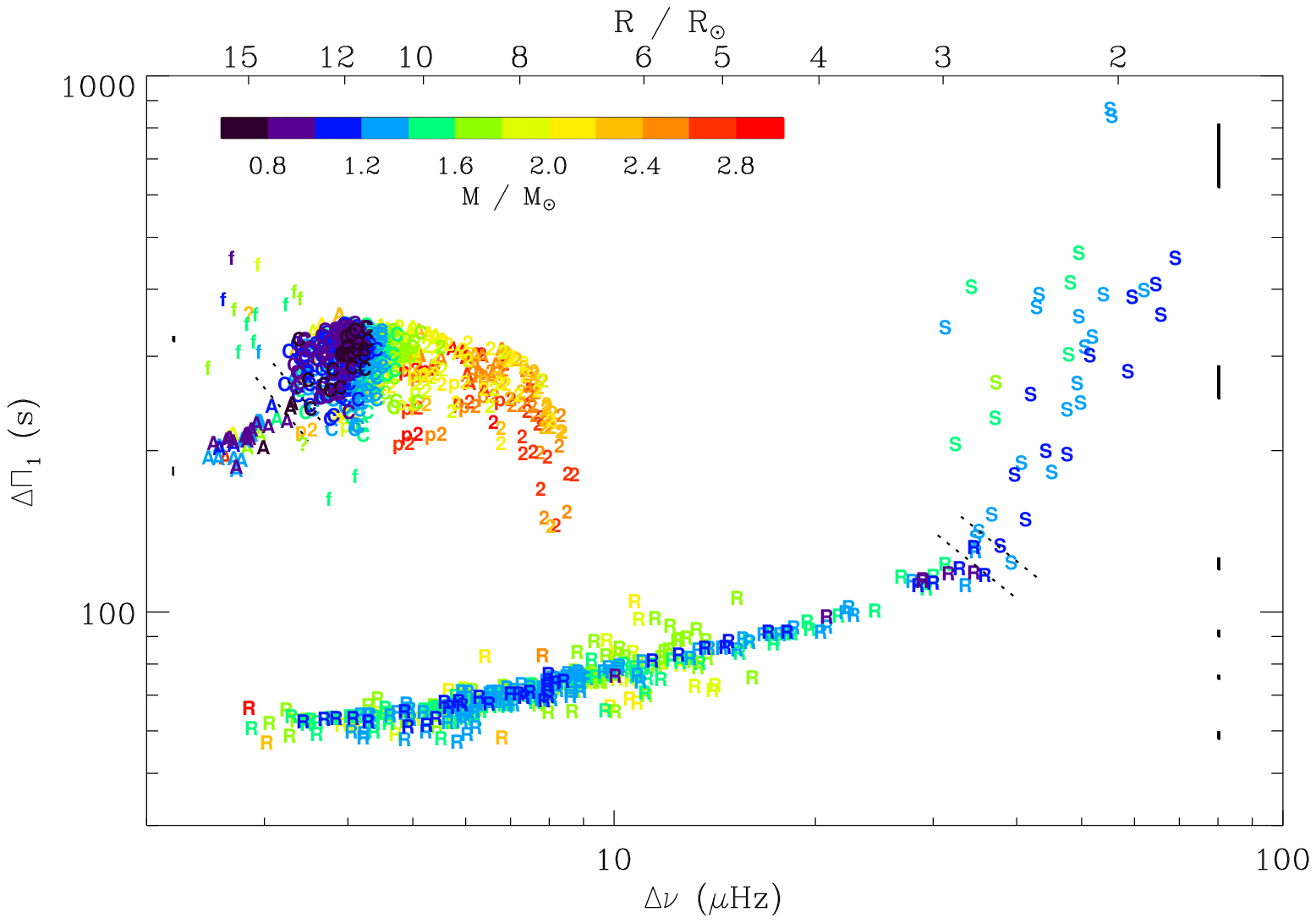}
\bigskip
\includegraphics[width=15.5cm]{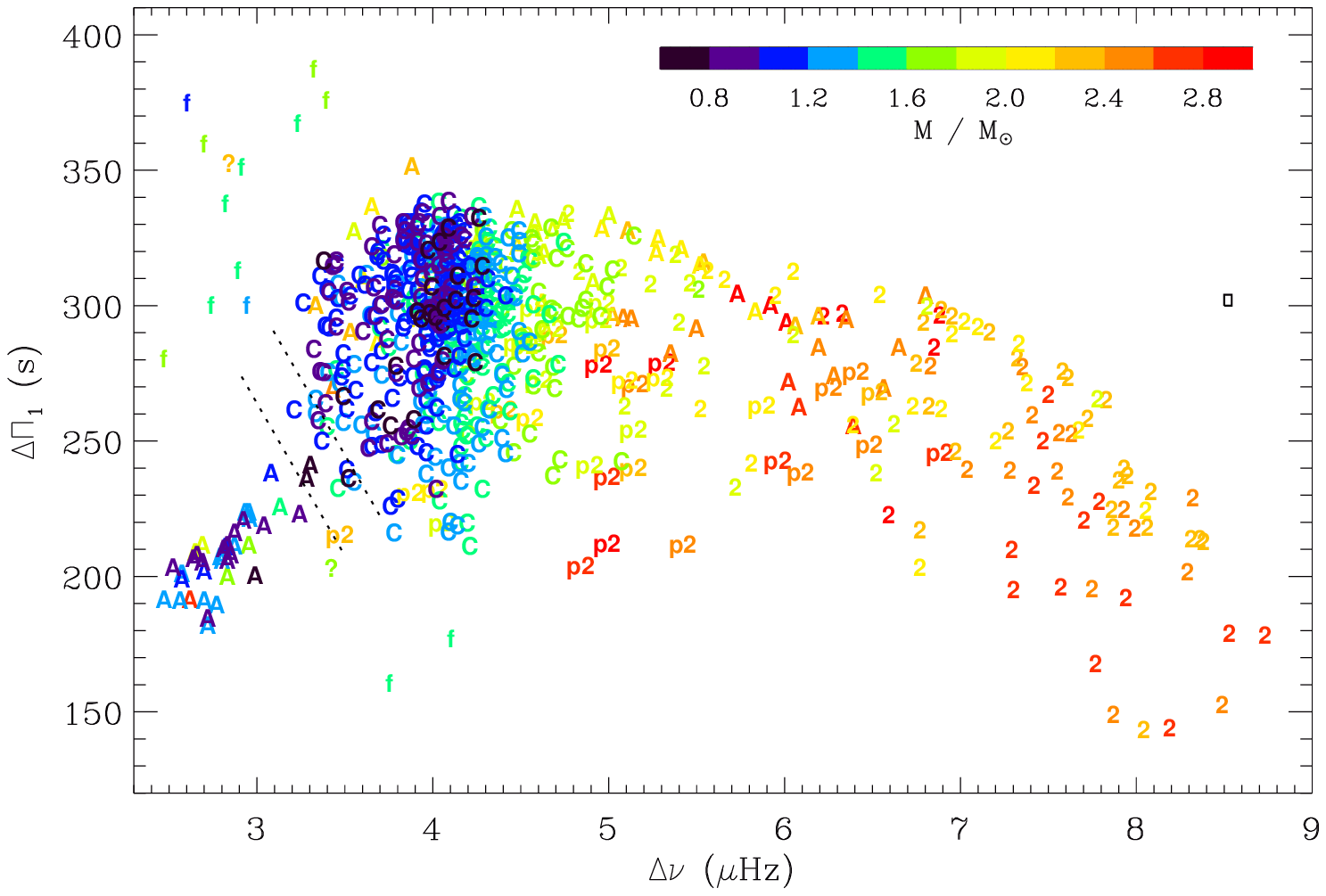}
\caption{Period spacing $\Tg$ as a function of the frequency
spacing $\Dnu$.
{\bf Top:} The seismic proxy for the stellar mass is indicated by
the color code. The evolutionary states are indicated by S
(subgiants), R (RGB), f (helium subflash stage), C (red clump), p2
(pre secondary clump), 2 (secondary clump), and A (stars leaving
the red clump moving toward the AGB). The error boxes on the right
side indicate the mean uncertainties, as a function of $\Tg$, for
stars on the RGB; for clump stars, uncertainties are indicated on
the left side. Dotted lines indicate the boundaries between
evolutionary stages.
{\bf Bottom:} Zoom in the red-clump region. Data used in this
figure are available at the CDS.
\label{fig_sismo}}
\end{figure*}

Examining the variation of $\Tg$ as a function of $\Dnu$ in a
seismic Hertzsprung-Russell (HR) diagram allows us to map stellar
evolution and to distinguish the late evolutionary stages
(Fig.~\ref{fig_sismo}). This $\HRsismo$ diagram provides richer
information than the classical HR diagram. Without seismic
data, determining precise evolutionary stages for field stars is
uncertain or impossible because of an observable
quantity that probes the innermost region of the stars and because
of the large uncertainties associated with the fundamental stellar
parameters (luminosity, effective temperature, and chemical
composition). Here, the density of stars in the $\HRsismo$ diagram
enables constructing seismic evolutionary tracks
(Fig.~\ref{fig_evolution}) directly derived from the mean
location of stars identified in mass ranges 0.2-$M_\odot$ wide.

\subsection{Subgiant stage\label{MSsub}}

A star leaves the main sequence and enters the subgiant stage
(stars {\bf S} in Fig.~\ref{fig_sismo}) when the hydrogen fuel is
exhausted in its core. In a star of mass lower than \Mmax\,
$M_\odot$, the observation of mixed modes indicates a dense radiative core and the beginning of the subgiant
phase, as is the case for the bright F-star Procyon
\citep{2010ApJ...713..935B}. The slow contraction of the core on a
thermal timescale implies a quasi-static adjustment, hence the
extension of the envelope. Accordingly, as a subgiant evolves, the
mean stellar density, and therefore the large separation,
decreases. The contraction of the radiative core also results in
a decreased period spacing. We also note a clear mass
dependence of the $\HRsismo$ relation (Fig.~\ref{fig_evolution}),
as predicted by stellar modeling \citep{2013ApJ...766..118M}.

\subsection{From subgiants to the red giant branch\label{fromsubtoRGB}}

\begin{figure}
\includegraphics[width=9cm]{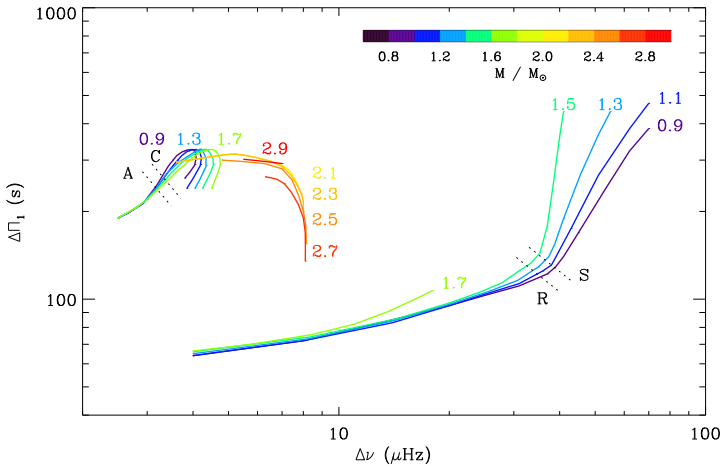}
\caption{Evolutionary tracks reconstructed from the seismic
observations for stellar masses in the $[0.9 - 2.9\,M_\odot$]
range. The $1.9\,M_\odot$ track is not shown because more
information is needed to define the limit between the clump and
secondary-clump stars. On the RGB, the dispersion due to the first
luminosity bump is too high to allow an unambiguous definition of
the evolutionary tracks for stellar masses above 1.9\,$M_\odot$.
Dotted lines indicates the boundaries between evolutionary stages.
\label{fig_evolution}}
\end{figure}

As stars evolve from subgiants onto the RGB, the increase of the
stellar radius induces the decrease of the large separation. We
note the convergence of the evolutionary paths in the seismic
$\HRsismo$ diagram. The properties of the stellar interior become
increasingly dominated by the physical conditions of the
quasi-isothermal degenerate helium core and its surrounding
hydrogen-burning shell \citep{2012sse..book.....K}. Accordingly,
the structural properties of the envelope are also related to the
core mass, which explains the degeneracy in the $\HRsismo$ diagram
for low-mass red giants (stars {\bf R} in Fig.~\ref{fig_sismo}).
Although the transition from the subgiant phase to the RGB can be
seen in the classical HR diagram, it is impossible to
unambiguously infer the evolutionary status of a given star from
its location in that diagram. In contrast, the evolution from the
subgiant to the red giant phase is clear in the $\HRsismo$
diagram: almost independent of the initial conditions (mass,
metallicity), all low-mass stars on the RGB with the same core
structure have the same mean density. The influence of metallicity
should be investigated to understand the higher dispersion
seen for stars more massive than \Mmax\,$M_\odot$; this is beyond
the scope of this work. We can summarize the change of regime with
an empirical criterion: a subgiant with a mass below
\Mmax\,$M_\odot$ starts climbing the RGB when
$(\Dnu/36.5\,\mu\hbox{Hz}) \,(\Tg / 126\,\hbox{s}) < 1$. This threshold is determined to better than 8\,\%, based on the
identification of the `elbow' in the evolutionary tracks. The
boundary and its uncertainties are indicated with dotted lines in
Figs.~\ref{fig_sismo} and \ref{fig_evolution}. Translated into a
stellar age, this uncertainty represents a very short event, much
shorter than 0.5\,\% of the evolution time on the main sequence.

\subsection{Structure of the red clump\label{clump}}

When the helium core of a low-mass star on the RGB reaches about
$0.47\,M_\odot$, runaway ignition in degenerate conditions
produces the helium flash, which very rapidly transports the star
from the tip of the RGB to the red clump
\citep{2002PASP..114..375S}. The highest mass a star can have to
undergo the helium flash is $1.9\, M_\odot$, with an uncertainty
of about 10\,\%. We did not take into account the scaling revision
proposed by \cite{2012MNRAS.419.2077M} for red-clump stars.
Red-clump stars (stars {\bf C} in Fig.~\ref{fig_sismo}) occupy a
small region of the $\HRsismo$ diagram, around 300\,s and
4.1\,$\mu$Hz \citep{2012A&A...540A.143M}. They have similar core
masses, hence similar luminosities, and are therefore used as
standard candles \citep{1998ApJ...494L.219P}. Seismic information
provides useful constraints for improving the structure of the red
clump, hence for improving distance measurements.

Models still have difficulties in reproducing the period spacing in
the red clump
\citep{2012ApJ...744L...6B,2013ApJ...766..118M,2013ApJ...765L..41S}.
In part this is due to an incorrect treatment of a chemical
discontinuity at the convective-core boundary in core-helium-burning models \citep{2014A&A...569A..63G}, resulting in
insufficient mixing in the core \citep{2013ASPC..479..435N}. More
importantly, the accuracy of the measurements of $\Dnu$ and $\Tg$
is high enough to track the evolution of the stars in the
helium-burning phase. Low-mass stars have lower $\Dnu$ than more
massive stars, hence lower mean density. This is in agreement with
the fact that the inner pressure is fixed by the hydrogen shell
that produces the largest part of the stellar luminosity. During
the first stage of helium burning, the core grows in mass and
expands, so that the envelope contracts: both $\Tg$ and $\Dnu$
increase. In a second stage, both decrease, due to a less
efficient energy production. This evolution is predicted by models
\citep{2012A&A...543A.108L,2013ApJ...766..118M,2013ApJ...765L..41S}.
Now, we can precisely quantify it (Fig.~\ref{fig_evolution}).

\subsection{Structure of the secondary clump}

In stars with masses above about $1.9\,M_\odot$, the ignition of
helium occurs gradually rather than in a flash because the core is
not fully degenerate
\citep{1999MNRAS.308..818G,2012ApJ...760...32H,2012MNRAS.419.2077M}.
Therefore, these secondary-clump stars (stars {\bf 2} in
Fig.~\ref{fig_sismo}) show a wider spread in the diagram
\citep{2012ApJ...744L...6B}: $\Tg$ decreases with increasing
stellar mass up to 2.7\,$M_\odot$, as does the mass of the helium
core at ignition. Then, for masses above 2.8\,$M_\odot$, $\Tg$
increases significantly with increasing stellar mass. This
behavior is expected from stellar modeling, which however often
fails at reproducing the mass corresponding to minimum $\Tg$
values \citep[e.g.,][]{2013ApJ...765L..41S}. This again emphasizes
the crucial role of the seismic HR diagram and the necessity of
accurately calibrating seismic scaling relations.

We defined the {\bf p2} status for each
mass range; this status
corresponds to possible progenitors of secondary-clump stars.
The mass-dependent threshold value between the two stages {\bf p2}
and {\bf 2}  was arbitrarily defined at $\Dnu$ lower than 25\,\%
of the median value in the secondary clump and $\Tg$ below the
mean value observed in the secondary clump for the considered mass
range. Progenitors of secondary-clump stars have a lower $\Dnu$ (a
higher luminosity) than the median stage in each mass interval,
and also a low $\Tg$ corresponding to an extended inner radiative
region. Comparison with modeling is necessary to confirm the
nature of these progenitors.

\subsection{From the red and secondary clumps to the asymptotic giant branch\label{AGB}}

A few stars appear in the vicinity of the clump, but with
significantly smaller period spacings.
They most probably correspond to stars in which the core is
contracting because helium becomes exhausted, leaving the main
region of the red clump and preparing to ascend the AGB
\citep{2012A&A...543A.108L,2012ApJ...757..190C,2013EPJWC..4303002M}.
Since a wide range of masses, including high masses, are present
on the same trajectory, we exclude the scenario that these stars
are entering the red clump. We empirically considered that a star
leaves the red clump and enters this stage, labeled with {\bf A}
in the $\HRsismo$ diagram, when its large separation is 15\,\%
below the mean value observed in the clump for stars with
similar masses. For low-mass stars, this occurs when
$(\Dnu/3.3\,\mu\hbox{Hz})^{1.5} \,(\Tg / 245\,\hbox{s}) < 1$. This threshold is determined to better than 6\,\%, based on the
narrowing of the evolutionary tracks. In the classical HR diagram,
such low-mass stars remain hidden in the red clump since they have
similar luminosity. This better characterization of the clump
stars is important for using them as accurate standard candles.

High-mass stars exiting the secondary clump can also be
identified. When core-helium burning becomes inefficient, $\Dnu$
decreases and $\Tg$ first increases. This behavior and the mass
criterion ensure a significant difference between the {\bf p2} and
{\bf A} stages. In a second step, $\Dnu$ and $\Tg$ vary as for
lower mass stars, but with a wider spread. This leads to a
reliable definition of the {\bf A} stage when $M\ge 1.9\,
M_\odot$, even with a limited set of stars.

\subsection{Helium flash}

Finally, we identified a small number of stars that clearly lie
outside the
evolutionary paths mentioned above. It is likely that these stars
have very recently undergone the helium flash (stars {\bf f} in
Fig.~\ref{fig_sismo}). At low $\Dnu$, we identified stars with an
unusually high period spacing, corresponding to a small inner
radiative region. For low-mass stars, this situation matches a
helium subflash \citep{2012ApJ...744L...6B}.  For higher mass, the
evolutionary stage cannot be determined among helium ignition or
blue-loop stage; two stars that may be in this stage are marked
with the symbol `{\bf ?}'. Stages between subflashes are hard to
detect because they have simultaneously small $\Tg$ and $\Dnu$
\citep{2012A&A...543A.108L}. We also identified stars with $\Tg$
just below the main clump, which certainly evolve toward the
clump with an almost fully ignited helium core.

\section{Conclusion}

Precise markers of stellar evolution of low-mass stars were derived in the $\HRsismo$ diagram. For each stellar mass interval,
evolutionary tracks were derived. All transitions between the
various stages of evolution, such as hydrogen-shell burning,
helium-core burning, and the end of helium burning in the core,
are marked by changes in the relationship between the frequency
and period spacings. For low-mass stars arriving on the RGB, the
period spacing varies with the frequency spacing because the core
and envelope structures are closely linked at that stage. Similar
variation is seen for core-helium-burning stars, but with a
mass-dependent relationship since nuclear burning in the core
removes the degeneracy of helium. When helium is exhausted, the
core is degenerate again, so that the relation between $\Tg$ and
$\Dnu$ is independent of mass, as on the RGB. A few outliers are
identified as stars starting the helium-burning stage in the
unstable contraction phase following the helium flash
\citep{2012ApJ...744L...6B}.

Comparison with modeling will help to link the phenomenological
threshold values  with evolution parameters such as the helium
fraction in the core, especially for secondary-clump stars, to
assess the {\bf p2} stage. This comparison is currently
impossible, since modeling has first to accurately reproduce the
observed tracks. Independent of this forthcoming analysis, we note
that there are fewer than 1\% of\ stars with ambiguous
identifications in the clump, which are identified as stars close
to the boundaries. Among these, we identified two high-mass stars
that might be in a blue-loop stage.

\begin{acknowledgements}
We thank the referee for his/her constructive comments. We
acknowledge the entire Kepler team, whose efforts made these
results possible. BM, KB, MJG, EM, MV, CB, and RS acknowledge
financial support from the Programme National de Physique
Stellaire (CNRS/INSU) and from the ANR program IDEE Interaction
Des \'Etoiles et des Exoplan\`etes. OB is supported by the Japan
society for promotion of science fellowship for research
(n$^\circ$ 25-13316). Funding for this Discovery mission was
provided by NASA's Science Mission Directorate. JDR acknowledges
support of the FWO Flanders under project O6260-G.0728.11 This
work partially used data analyzed under the NASA grant
NNX12AE17GPGB and under the European Communityfs Seventh
Framework Program grant (FP7/2007--2013)/ERC grant agreement n.
PROSPERITY.
\end{acknowledgements}

\bibliographystyle{aa} 

\end{document}